%% file: mnras_template.tex
\documentclass[fleqn,usenatbib]{mnras}

\usepackage{newtxtext,newtxmath}

\usepackage[T1]{fontenc}
\usepackage{ae,aecompl}

\usepackage{ulem}
\usepackage{enumitem}

\usepackage{graphicx}	
\usepackage{amsmath}	
\usepackage{amssymb}	
\usepackage[usenames,dvipsnames]{xcolor}
\usepackage{color}
\usepackage{bbding}
\usepackage{tikz}
\usepackage[switch,modulo]{lineno}
\usepackage{comment}
\usepackage{dcolumn}

\newcommand{\fellwalker}{{\texttt{Fellwalker}}}
\newcommand{\kms}{km s$^{-1}$}

\title[Peering into Carina's Pillars with ALMA]{Carina's Pillars of Destruction: the view from ALMA}

\author[P.D. Klaassen et al.]{
P. D. Klaassen,$^{1}$\thanks{E-mail: pamela.klaassen@stfc.ac.uk}
M. R. Reiter,$^{1}$
A.F. McLeod,$^{2,3}$
J.C. Mottram,$^{4}$\newauthor
J.E. Dale,$^{5}$
and M. Gritschneder$^{6}$
\\
$^{1}$UK Astronomy Technology Centre, Royal Observatory Edinburgh, Blackford Hill, Edinburgh, EH9 3HJ, UK\\
$^{2}$Department of Astronomy, University of California Berkeley, Berkeley, CA 94720, USA\\
$^{3}$Department of Physics and Astronomy, Texas Tech University, PO Box 41051, Lubbock, TX 79409, USA\\
$^{4}$Max Planck Institute for Astronomy, K\"onigstuhl 17, 69117 Heidelberg, Germany\\
$^{5}$Centre for Astrophysics Research, University of Hertfordshire, College Lane, Hatfield, AL10 9AB, UK\\
$^{6}$Universit\"ats-Sternwarte, Ludwig-Maximilians-Universit\"at Munchen, Scheinerstr. 1, 81679 Munich, Germany }

\date{Accepted XXX. Received YYY; in original form ZZZ}

\pubyear{2018}

\begin{document}
\label{firstpage}
\pagerange{\pageref{firstpage}--\pageref{lastpage}}
\maketitle

\begin{abstract}
Forming high-mass stars have a significant effect on their natal environment. Their feedback pathways, including winds, outflows, and ionising radiation, shape the evolution of their surroundings which impacts the formation of the next generation of stars.  They create or reveal dense pillars of gas and dust towards the edges of the cavities they clear.  They are modelled in feedback simulations, and the sizes and shapes of the pillars produced are consistent with those observed. However, these models predict measurably different kinematics which provides testable discriminants.  Here we present the first ALMA Compact Array (ACA) survey of 13 pillars in Carina, observed in $^{12}$CO, $^{13}$CO and C$^{18}$O J=2-1, and the 230 GHz continuum.  The pillars in this survey were chosen to cover a wide range in properties relating to the amount and direction of incident radiation, proximity to nearby irradiating clusters and cloud rims, and whether they are detached from the cloud.  With these data, we are able to discriminate between models.  We generally find pillar velocity dispersions of $<$ 1 \kms{} and that the outer few layers of molecular emission in these pillars  show no significant offsets from each other, suggesting little bulk internal motions within the pillars. There are instances where the pillars are offset in velocity from their parental cloud rim, and some with no offset, hinting at a stochastic development of these motions.  

\end{abstract}




\section{Introduction}
Out of the chaos involved in the formation of high-mass stars come the beautiful pillar structures seen at the rims of the ionised bubbles produced when the stars feed radiation back into their environment \citep[e.g.,][]{Hester1996,White1999,Westmoquette2013,Hartigan15,Schneider2016}. 
The structure and dynamics of these pillars can tell us about the past, present and future star-formation in the region, and are naturally produced in simulations of ionisation feedback \citep[e.g.,][]{Gritschneder10,Mackey2010,Tremblin2012}.  When young massive \citep[M $\gtrsim$ 8 M$_\odot$][]{Zinnecker07} stars reach the main sequence, their intense radiation begins to ionise and destroy the parent molecular cloud. In combination with the original cloud structure and properties, the way this feedback proceeds influences star formation efficiency on a range of scales \citep[see, for instance][]{Mellema06}. Most stars form in a hierarchically-clustered fashion \citep{Krumholz2018StarTime}, and low-mass stars are likely to experience the effects of this feedback. For example, feedback can truncate and accelerate the destruction of proto-planetary disks \citep[e.g.]{Adams04,Eisner2018}, and thus influence planet formation \citep[see for instance,][]{Throop2005,Nicholson2019}.  In some cases it can also sweep up or induce material to form a subsequent generation of stars \citep[see, for instance][]{Liu15,Bisbas11}. Furthermore, it is the principal process by which molecular clouds are destroyed \citep{Matzner02}, and how quickly this occurs dictates the impact and propagation of material from supernovae into the cloud and surrounding intertstellar medium  \citep[ISM;][]{Walch14,Iffrig2015}. The progress of this ionising feedback therefore has a profound effect on the star formation cycle from cloud core to Galactic scales (e.g. 0.1pc - kpc).

The parameter space of initial conditions and methods for forming pillars has been comprehensively studied by various theoretical models over recent years. These include external radiation hitting Bonnor-Ebert spheres \cite{Gritschneder09}, turbulent media \cite{Gritschneder10}, fractal clouds \cite{Walch12,Walch13}, or radiation from a cluster inside a molecular cloud \cite{Dale12II}.   In these types of simulations, pillars arise either from the lower density ambient material being preferentially swept away, or through instability growth.

The key measurable constraints common to the models described above, which then relate directly back to the effects of ionised feedback, are:

\begin{itemize}
\item the presence (or lack thereof) of internal and surface motions
\item the internal and surface velocity dispersions, and how they vary within and along the pillar
\item velocity offsets or gradients between the pillar and the parent cloud, so called `ordered-flows' of material
\item the presence, evolutionary stage, and spacing of any cores/protostars within the pillar 
\end{itemize}

The individual models result in  different star formation efficiencies and timescales due to differences in the density contrast between the pillars and the surrounding medium, different progression speeds of the ionisation front, as well as whether cores were already present or were induced to collapse by the ionisation-driven shock. These different initial conditions and formation mechanisms result in measurable differences in the gas kinematics within, and around, the pillars.

Spatially and spectrally resolved observations of gas kinematics in pillars are therefore an excellent way to constrain model predictions of the results of ionising feedback on molecular clouds. With spatially and spectrally resolved observations, we are able to directly compare similar structures between observations and simulations. If the pillars are not resolved, we cannot quantify their internal dynamics - motions which can differ between models, as described above.

Despite the importance of constraining cloud destruction, there has so far only been one study that has both spatially and spectrally resolved the kinematics of a single large pillar \citep{Klaassen14}, although see \citet{Rebolledo16} and \citet{Dawson11} for spectrally resolved studies of more than one pillar.  The study of \citeauthor{Klaassen14}, while providing a good proof of concept, could not constrain the general processes underlying ionised feedback and cloud disruption in general.  To do that, a moderately high resolution survey of multiple pillars in a single environment is required. 

In this paper, we present an overview of the first ALMA Compact Array (ACA) survey of this kind, focusing on CO and its isotopologues in 13 pillar structures in Carina, with a comparison to models. Carina was chosen for this study because of its relative proximity \citep[2.5 kpc][]{Povich19}, and wealth of high mass stars \citep[see, for instance, ][and others]{Smith06b,Gagne11,Alexander16} that are shaping the region. Many of the hundreds of OB-type stars in Carina lie outside the main stellar clusters. This leads the UV field/ionising photon flux in Carina to vary between modest and extreme, with irradiation near the central clusters approaching or exceeding that measured in famous photon-dominated regions \citep[G$_0 \gtrsim 10^4$][]{Brooks03,Wu18} like the Orion Bar and M17 \citep[see, for instance][respectively]{Goicoechea16,Sheffer13}. 

The overarching goal of this project is to quantify the kinematics of pillars in Carina, to determine whether the simulations are accurately predicting the effects of feedback on the local environment.  With these results, theorists can determine which properties their models are capturing in a realistic manner, and modify their future simulations accordingly.

\begin{figure*}
\includegraphics[width=\textwidth]{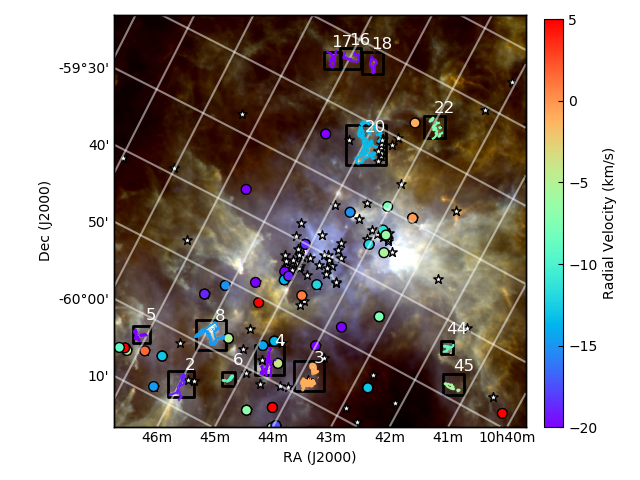}
\caption{\textit{Herschel} RGB image of the region surrounding eta Carina (250, 160 and 70 $\mu$m, respectively). The circles show the positions of the O stars from Table 4 of \citet{Hanes18} and Table 2 of \citet{Kiminki18}. The colours of the star markers corresponds to their radial velocities of those stars.  The O stars without radial velocity measurements \citep[as presented in][]{Alexander16} are shown with black stars. The black boxes highlight the regions imaged with ALMA, and the contours within show the 10$\sigma$ contour of the CO integrated intensity, where the colour represents the mean velocity of the gas on the same colour scale as the stars. The numbers of the pillars as listed in Table \ref{tab:centers} are shown in white above each pillar box. Given a distance to Carina of 2.5 kpc, the linear scale of the map is $\sim$45 pc on a side.}
\label{fig:context}
\end{figure*}

The observed pillars were chosen to sample a variety of morphologies and ambient properties. From the \textit{Herschel} images of the region (See Figure \ref{fig:context}), some of the pillars are broad, while others are narrow. Some are detached from the edges of the cavities (the so called `rims') while others appear as small fingers protruding from larger rims. The proximity of O and B stars to the pillars, and the numbers of these stars, are also highly varied. Taken together, these pillars can be viewed as representative of the general population of pillars in Carina. The pillars in this survey are large structures ($\gtrsim$ 0.1 pc) that often, but not always, point toward clusters of ionising stars in the region. \citet{McLeod16} examined the resulting ionisation fronts at the tips of some of these pillars  and found a clear correlation between the incident ionising flux and both the ionisation front density and the photo-evaporation rate off the cloud surface.

We compare these observed pillars to the feedback generated pillars in the models of \citep{Gritschneder09,Gritschneder10,Dale12II} to determine whether we can distinguish between the models, and therefore aid in their future refinement.

 Many of the modelling properties of these studies are quite similar.  Both use smooth particle hydrodynamics (SPH) codes with 1-2$\times10^6$ particles, have initial temperatures less than $\sim$ 100 K (generally closer to 10 K), an initial turbulent velocity fields that are supersonic (with the gas generally moving slower than Mach 20). 

In terms of initial conditions, the key differences between the models is the treatment of turbulent decay and the initial bulk density of the gas.  In \citet{Gritschneder09,Gritschneder10}, they use $E\propto k^{-2}$, while in \citet{Dale12II}, they use a Kolmogorov decay, with $E\propto k^{-5/3}$.

The volumes under consideration in the models are different, with \citeauthor{Gritschneder10} using boxes with side lengths of 2-8 pc, while \citeauthor{Dale12II} used boxes with sides of a few to 100 pc. However, the bulk densities are within a factor of 10 of each other, and are consistent with the bulk density in Carina \citep[See the discussion in ][]{Yonekura05}.

Below, in Section \ref{sec:observations}, we present an overview of our observations, and in Section \ref{sec:analysis} we present our initial findings from these observations. In Section \ref{sec:discussion} we compare these results to the predictions of feedback models, and discuss the implications of the velocity structures in these pillars. We summarise and conclude in Section \ref{sec:conclusions}

\section{Observations}
\label{sec:observations}

\input{obs_params.tab}

The pillars sample the `Southern Pillars', `Northern Pillars', and the `Southwestern Pillars' complexes as defined in \citet{Hartigan15}, and in the following we use their pillar numbering scheme to refer to the pillars observed in this work.  Those in the southern region (Pillars 2, 3, 4, 5, 6, and 8) are all part of the region shown by \citet{Smith10_spitzer} to have a higher than average young stellar object (YSO) population, as identified by \textit{Spitzer} IRAC observations, than the rest of the pillar regions in Carina.

We present ALMA Compact Array (ACA) and Total Power (TP) observations of CO, $^{13}$CO and C$^{18}$O (J=2-1) taken in October and November 2016 (ACA) and with the TP data primarily taken between October 19 and 21, with the exception of the TP data for Pillar 3, which was taken on 3 March, 2017. These data form ALMA project \#2016.1.00101.S.  All data were pipeline reduced in CASA 4.7 \citep{CASA}. Because of the bug in CASA 4.7 with respect to creating mosaics, the ACA 7m data were re-mosaiced (re-imaged) in CASA 5.4. No further self-calibration was applied because the continuum emission was not well detected in all mosaics. The time on target, calibrators used, and precipitable water vapour for each mosaic are presented in Table \ref{tab:obs_params}, with map centers presented in Table \ref{tab:centers}.

\newcolumntype{d}[1]{D{.}{.}{#1} }
\begin{table}
\centering
\caption{Centres of Observed Mosaics.}
\begin{tabular}{cd{1}d{3}}
\hline \hline
Pillar & \multicolumn{1}{c}{{Right Ascension}} & \multicolumn{1}{c}{{Declination}} \\
&  \multicolumn{1}{c}{{(h:m:s)}} & \multicolumn{1}{c}{{(d:m:s)}}\\
\hline
2 & 10:45:58.60 & -60:06:36.331 \\
3 & 10:43:50.17 & -59:56:48.825 \\
4 & 10:44:39.99 & -59:57:26.020 \\
5 & 10:47:06.20 & -60:02:45.529 \\
6 & 10:45:11.84 & -60:02:40.359 \\
8 & 10:45:53.81 & -59:58:09.603 \\
16 & 10:46:16.50 & -59:14:32.486 \\
17 & 10:45:59.27 & -59:13:02.708 \\
18 & 10:45:34.51 & -59:12:06.091 \\
20 & 10:44:57.26 & -59:23:14.049 \\
22 & 10:43:58.05 & -59:16:09.442 \\
44 & 10:41:44.99 & -59:43:38.046 \\
45 & 10:41:18.20 & -59:47:52.800 \\
\hline
\end{tabular}
\label{tab:centers}
\end{table}

\begin{table}
\centering
\caption{Observed Species, synthesised beams and noise levels per channel. The CO isotopologues show the values for the feathered data. The rest are for ACA only.}
\begin{tabular}{rr@{$\times$}rrrr}
\hline \hline
Species & \multicolumn{3}{c}{Synth. Beam} & Noise & Ch. Width\\
  (Trans.)  & ($''$)& ($''$) & ($^\circ$) & (Jy/beam) & (km/s)\\
 \hline
 CO (J=2-1) & 6.59 & 5.03 & 43 & 0.60&0.079\\
 $^{13}$CO (J=2-1) & 7.06 & 5.10 & 85.7 & 0.66& 0.083 \\
 C$^{18}$O (J=2-1) & 7.16 & 5.18 & 83.3 &0.49 & 0.083\\
 $^{13}$CS (J=5-4) & 6.41 &4.61 & 77.8 & 0.41 & 0.158\\
 SiO (J=5-4) & 6.64 & 5.09 &  81.9 & 0.52& 0.158\\
 \hline
 Continuum & 6.25&4.71 & 75.8 & 6$\times 10^{-5}$& 1.87 GHz\\
 \hline\hline
\end{tabular}
\label{tab:beam_sens}
\end{table}

Along with CO and its isotopologues, we had additional spectral windows set to detect the 233 GHz continuum emission, as well as $^{13}$CS (J=5-4) and SiO (J=5-4). Neither of these last two species were detected. The synthesised beams and sensitivity limits (per bin) are presented in Table \ref{tab:beam_sens}, with the note that the CO isotopologue beams and sensitivities are for the final combined (feathered) data, while the others represent ACA only data.

\subsection{Feathering  the ACA and Total Power Data}

\begin{figure*}
\includegraphics[width=\textwidth]{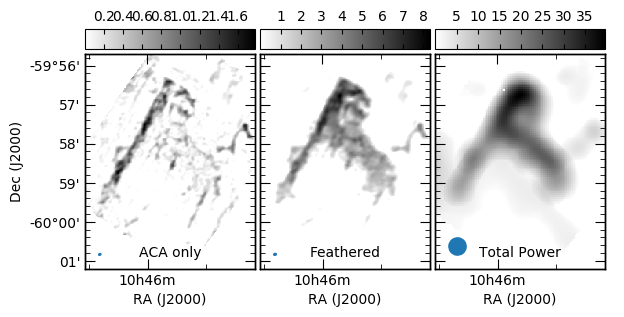}
\caption{Integrated intensity of the CO (J=2-1) emission from Pillar 8. {\bf Left:} ACA only data, {\bf Centre:} Feathered ACA and Total Power (TP) data, {\bf Right:} TP data regridded to the resolution of the ACA data. The units of the colourbars are Jy/beam.}
\label{fig:feathering}
\end{figure*}

Despite using the ACA in stand alone mode, our interferometric data suffered from spatial filtering of the large scale emission.  To capture the largest scale structure in our mosaics, we additionally obtained total power (TP) datasets to complement our 7m data. For the spectral windows with CO (and isotopologue) detections, we combined these datasets using the \verb+CASA+ task \verb+feather+.  To feather the data, both datasets were first imaged on the same velocity grid, and the TP data was spatially regridded using the ACA data as a template. The TP data was then converted to the same flux scale as the ACA data, and the two data cubes were feathered together.  Figure \ref{fig:feathering} shows the ACA only, feathered and TP only datasets for Pillar 8 to highlight how the combination of ACA and TP data improves the sensitivity to large scale structure in the CO emission. 

\subsection{Systematic Uncertainties in ALMA data}

 In addition to the formal uncertainties on the data, such as the rms noise of each map (see Table \ref{tab:beam_sens}) and the velocity sampling of our spectral channels, we describe here the other, more systematic sources of uncertainty in our observations. We describe the effects of these systematics on our derivations, their ranges, and, where appropriate, the reasons for choosing the reference values we have.

The first of these systematics, and arguably the easiest to quantify, is the flux calibration of our observations.  With each delivered dataset, both the flux and flux uncertainty of each flux calibrator was provided.  While these uncertainties varied slightly between observing dates and calibrators, in all cases the flux calibration uncertainty was less than 6\%. We use this upper envelope in our mass and column density calculations presented in Section \ref{sec:MN}.

In addition to this, there is uncertainty in the distance to Carina. The canonical distance has generally been accepted as 2.3 kpc \citep{Smith06b}, However more recent distances derived from {\it Gaia} measurements place Carina at distances of between 2.5 and 2.6 kpc \citep{Kuhn19,Davidson18,Povich19}.  We adopt the 2.5 kpc distance estimate of \citeauthor{Povich19}, because of their more robust outlier rejection algorithms.  We note that the distance uncertainty would propogate linearly through our clump separation calculations in Section \ref{sec:jeans}, while the column densities and masses derived in Section \ref{sec:MN} vary with distance squared.

A third source of systematic uncertainty is in the abundance ratio of CO and its isotopologues with respect to H$_2$. In low extinction regions, the abundance of CO can be suppressed to levels on the order of $5\times10^{-6}$ \citep{Goldsmith08,vDB88}, although it does still exist at these low extinction values \citep{Liszt17}.  Self shielding is less critical for the other CO isotopologues \citep{Visser09}, so their abundances are less likely to be suppressed. Because much of our CO emission comes from regions with column densities greater than the threshold for star formation \citep[e.g. $2 \times 10^{21}$ cm$^{-2}$,][ see the white contour on the {\it Spitzer} emission in Figures \ref{fig:pillar2}-\ref{fig:pillar45}]{Clark14}, we use the canonical CO abundance of $10^{-4}$ presented in \citet{Frerking82}, which is consistent with the more recent estimations that take galactocentric radii into account (i.e. $9.7\times10^{-5}$ using equation 6 of \citet{Pon16} and a galactocentric radius for Carina of 8.3 kpc). \citeauthor{Frerking82} point out that uncertainties in the CO abundance can be up to a factor of 3, which sould scale linearly with our derived column densities and masses.

\subsection{Supplementary MUSE data}

Where there is survey overlap, we compare out ALMA observations to those obtained with MUSE in the study of \citet{McLeod16}.  In that study, the surface properties of pillars in the Carina Nebula Complex were derived, including velocity maps of the ionisation fronts.

\section{Results}
\label{sec:analysis}

Figure \ref{fig:context} shows the regions observed in this study. The RGB colourscale shows the {\it Herschel} 250, 160 and 70 $\mu$m emission, with the black boxes highlighting the regions mapped with the ACA in this study. The single (coloured) contour in each black box corresponds to the 10$\sigma$ intensity limit on the CO integrated intensity map. The colours of these contours correspond to the mean velocity of the CO gas in that region (i.e. the mean of the intensity weighted velocity map of the pillar).  Plotted on the same colour scale are the radial velocities of the  surrounding O stars taken from the 2014 epoch measurements of \citet{Hanes18}, and Table 2 of \citet{Kiminki18}, where the colour reflects the mean measured radial velocity for each star.  The radial velocities of the stars are plotted on the same colour scale as the CO to highlight which ones are more likely to be influencing the observed molecular gas (i.e. similar colours could indicate a stronger connection between ionising source and pillar). Where there are discrepancies between the velocities reported in \citeauthor{Hanes18} and \citeauthor{Kiminki18}, we have used the \citeauthor{Kiminki18} values, because they have accounted for the binary orbits of the stars. The positions of the Carina O stars without radial velocity measurements, taken from \citet{Alexander16}, are plotted as black stars for reference.

Figures \ref{fig:pillar2} - \ref{fig:pillar45} show  the {\it Spitzer} archive 8 $\mu$m emission from each imaged region in the upper left panel, with the RGB images of the CO, $^{13}$CO and C$^{18}$O emission (respectively) in each pillar in the upper right.  Together these images show the shape of the warm dust ({\it Spitzer} image), and the molecular gas (CO RGB image) in the observed pillars. The bottom left panels of these figures shows the moment one (intensity weighted velocity) maps of the $^{12}$CO in each region, to highlight the bulk gas velocities of each pillar. The spectrum in the bottom right panel of each figure shows the $^{12}$CO (red), $^{13}$CO (green), and C$^{18}$O (blue) spectrum of each map averaged over the entirety of the region observed.  Overplotted on the {\it Spitzer} images in the top left panel of each figure is a contour of the CO integrated intensity map set to 10.4 Jy beam$^{-1}$ \kms{}. This contour corresponds to an H$_2$ column density of $2\times10^{21}$ cm$^{-2}$, and the threshold for star formation described in \citet{Clark14}.

Below we present the bulk properties of the observed pillars. From our CO and isotopologue observations of these 13 pillars, we have been able to derive an optical depth corrected molecular column density, from which we can derive individual pillar masses.  Where each isotopologue is detected, we can determine mean velocities and velocity gradients across each pillar. With these pillar properties, and the velocity dispersions of the gas, we can determine the virial stability of the gas and whether individual pillars are moving with respect to the rims they are attached to (when applicable).

For all calculations involving velocities, we clipped the velocity ranges of the cubes to contain only emission between -30 \kms{} and + 10 \kms{}, which includes the local standard of rest velocity of Carina \citep[$\sim$ -20 \kms{} suggested in ][]{Rebolledo16}. This is because for some regions (e.g. Pillar 3) there appear to be other CO emission components along the line of sight, which are unlikely to be from the pillar under analysis (i.e. likely to be foreground or background contamination).

\subsection{Pillar Column Density and Mass}
\label{sec:MN}

Because we observed multiple CO isotopologues in the same transition, we are able to calculate the optical depth of the CO emission. This was done using equation 1 of \citet{Choi93}.  For every channel in each map where the $^{13}$CO emission was above six sigma we calculated the optical depth of the CO primarily using the $^{13}$CO as the optically thin transition, and an abundance ratio of [CO]/[$^{13}$CO] = 60, the abundance ratio expected at the distance of Carina \citep{Rebolledo16}. A cutoff of six sigma was used in order to ensure a robust detection in all isotopologues used in the calculation. The reason we used the $^{13}$CO emission for our clipping mask to ensure that there was at least a minimal $^{13}$CO detection where we are doing the calculations.   For most positions in each map, the optical depth of the CO emission was less than 60, which indicates that the $^{13}$CO was optically thin.  When the calculated optical depth was greater than 60, generally C$^{18}$O was detected, and the emission and abundance ratios between the CO and C$^{18}$O was used instead \citep[560,][]{WR94}.

When calculating the column density in each pixel of each map, the CO emission was scaled by the optical depth following:

\begin{equation}
N_{CO} = \frac{Q_{rot}}{g_u}e^{E_u/kT_{ex}} * \frac{8\pi k_B \nu^2}{A_{ij}hc^3}Tdv * \frac{\tau_{12}}{1-e^{-\tau_{12}}}
\end{equation}

\noindent where the first portion of the equation scales the J=2-1 emission to the total CO emission given an excitation temperature of 30 K, and the final portion is the correction for optical depth \citep[See, for example][]{Mangum15}. Using a CO abundance of 10$^{-4}$ with respect to H$_2$, we find mean pillar column densities of order 1$\times10^{20}$ cm$^{-2}$, roughly half the column density required for star formation, with peak column densities 2-3 orders of magnitude higher (see Table \ref{tab:MN}). 

From the column density at each position, and in each velocity channel, we can determine the total gas mass for each pillar using the area of each pixel.  From the column density and area of each pixel, we calculate the number of molecules, and from that, assuming a mean molecular weight of 2.8 times the mass of a hydrogen atom \citep[see, for instance ][]{Kauffmann08}, we calculate the mass in each pixel, which is then summed across each map (above the six $\sigma$ threshold) to give the total gas mass in each pillar.  

Overall, we find a wide variety in pillar masses, between some 10s and a few 100 of M$_\odot$. This large range in masses is in part due to the variable sizes of the pillars, and portions thereof that we imaged (see Figure \ref{fig:context}).

\begin{table}
\centering
\caption{CO Derived Masses and Column Densities for each Pillar under the assumption of T = 30 K, and d= 2.5 kpc. }
\begin{tabular}{cr@{$\pm$}lr@{$\pm$}lr@{$\pm$}lr}
\hline \hline
Pillar & \multicolumn{2}{c}{Mass}& \multicolumn{2}{c}{Peak N} & \multicolumn{2}{c}{Mean N} & Velocity \\
 & \multicolumn{2}{c}{$\mathrm{(M_{\odot})}$}& \multicolumn{2}{c}{$\mathrm{(10^{21} cm^{-2})}$}& \multicolumn{2}{c}{$\mathrm{(10^{19} cm^{-2})}$} & (km/s)\\
\hline
2 & 501 & 30 & 8.3 & 0.7 & 51 & 3 & -18.83 \\
3 & 187 & 11 & 6.5 & 0.5 & 45 & 3 & -1.04 \\
4 & 203 & 12 & 2.8 & 0.3 & 50 & 3 & -26.02 \\
5 & 22 & 1 & 2.2 & 0.2 & 42 & 3 & -22.38 \\
6 & 48 & 3 & 6.1 & 0.4 & 46 & 3 & -9.61 \\
8 & 1386 & 83 & 12.4 & 1.2 & 57 & 3 & -14.77 \\
16 & 124 & 8 & 2.9 & 0.3 & 49 & 3 & -19.53 \\
17 & 56 & 3 & 1.3 & 0.1 & 35 & 2 & -20.12 \\
18 & 124 & 7 & 4.0 & 0.4 & 63 & 4 & -19.71 \\
20 & 449 & 27 & 12.9 & 1.0 & 88 & 5 & -13.41 \\
22 & 89 & 5 & 1.8 & 0.2 & 39 & 2 & -7.51 \\
44 & 101 & 6 & 4.3 & 0.4 & 86 & 5 & -8.91 \\
45 & 25 & 1 & 1.0 & 0.1 & 21 & 1 & -5.48 \\

\hline
\end{tabular}
\label{tab:MN}
\end{table}

\subsection{Separations Between Intensity Peaks}
\label{sec:jeans}

From the integrated intensity (moment zero) maps of the CO emission, we identified the `clumpiness' of the molecular gas in each region using a \fellwalker{} analysis \citep{fellwalker}.  Generally, this type of analysis is done on continuum emission rather than molecular (i.e. on cores rather than molecular intensity peaks), however the dust continuum in the observed pillars was poorly detected, with faint detections in only 7  pillars. Conversely, the CO emission is fairly contiguous for most pillars (with the exception of Pillar 3 which was very flocculent, despite the CO being well detected), and was therefore used as a proxy.  We present these results to estimate locations of likely over-densities.

\begin{figure}
\includegraphics[width=\columnwidth]{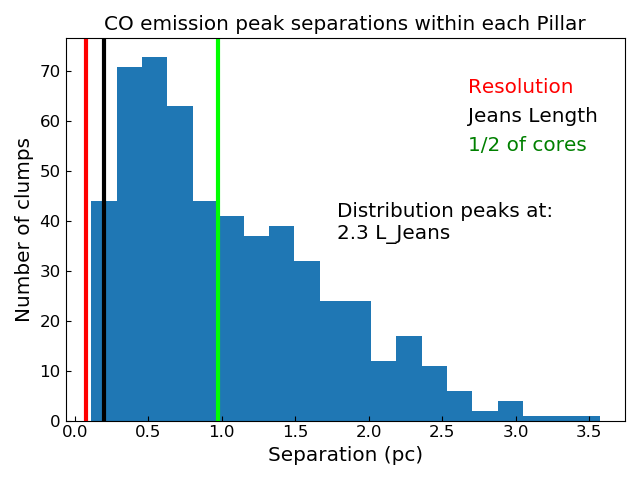}
\caption{Histogram of molecular emission peak separations in the observed pillars. The centers of the peaks were found with a \fellwalker{} analysis, and the distances between them found using a minimum spanning tree algorithm. The red line indicates the resolution of our observations, the black line the local Jeans length of 30 K gas, and the green line the dividing line between clustered and dispersed peaks.}
\label{fig:core_seps}
\end{figure}

From the \fellwalker{} analysis, we determined the peak positions of each CO emission peak, and, using a minimum spanning tree analysis, determined the separations of these peaks in all pillars. The results of this minimum spanning tree analysis are shown in Figure \ref{fig:core_seps}.  Also plotted here are, in red, the resolution of our observations, and in black, the Jeans length of 30 K gas.  Lowering the temperature or density would push the black line to larger sizes.  The green line highlights the median of CO core separations in the observed pillars.  It shows that the median is well above both the observing resolution and the `current' Jeans length for these pillars. Interestingly, this length scale corresponds to the Jeans length if the temperature were instead 10 K, which is more like the temperatures expected in cold cores as would have been the case when these pillars were starting to form.

\subsection{Pillar Velocities}
\label{sec:velocities}

The mean radial velocity of each pillar is given in the last column of Table \ref{tab:MN}, and shows a distribution of velocities most of which are near to the local standard of rest velocity of -20 km s$^{-1}$ expected for Carina \citep{Rebolledo16}.  The bottom right panels in Figures \ref{fig:pillar2} - \ref{fig:pillar45} show the intensity weighted velocity (moment 1) maps of the CO emission in each pillar, with the fourth showing the spectra of each isotopologue averaged over the entire map, with the C$^{18}$O emission scaled up by a factor of 5 to be visible. 

\begin{figure}
\includegraphics[width=\columnwidth]{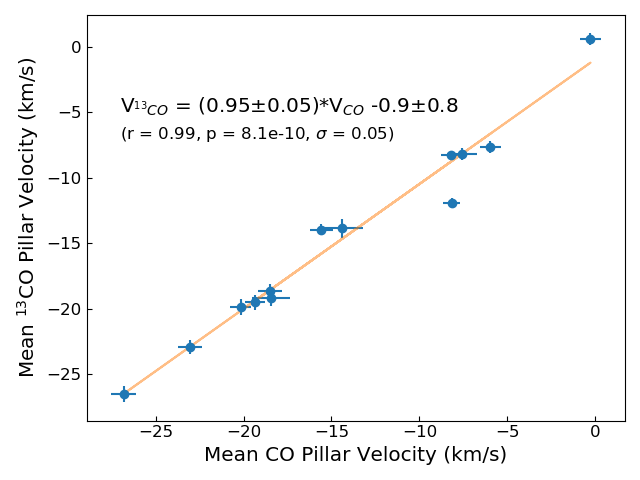}
\caption{Measure of the mean motion of the CO in each pillar compared to that of $^{13}$CO. The plotted error bars represent the mean of the velocity dispersions across each map in the respective isotopologues, because the formal uncertainties on the velocities (1/2 the channel width, or $\sim$\,0.04 km/s) would not be visible on these velocity scales.}

\label{fig:internal_motion}
\end{figure}

Noticeable in these moment one maps is that many of the pillars either have velocity gradients across the width of the pillar (e.g. Pillars 2, and 17), or that the pillar has a significantly different velocity (> 0.3 km s$^{-1}$) from the cloud rims that they are associated with (e.g. Pillars 5, 22, 44, and to a certain extent 8).  Interestingly though, comparing the mean velocity of the CO in each pillar to the $^{13}$CO shows no significant differences between the regions probed by these isotopologues: the surface motions as traced by CO are consistent with the interior motions as traced by $^{13}$CO.  How well the pillar surface and internal motions agree can be seen in Figure \ref{fig:internal_motion} where we plot the average CO velocity against the average $^{13}$CO velocity.  The y-intercept in Figure \ref{fig:internal_motion} also shows a minimal systematic velocity offset of 0.9$\pm$0.8 \kms{} between the CO and $^{13}$CO. The C$^{18}$O emission was only detected in the densest cores within the pillars, and as such, its mean velocity cannot be used as a deeper probe of the same information.

Pillars 5, 22, and 44 show distinct motions of the pillars with respect to the cloud edges they are associated with.  The emission patterns in Pillars 22 and 44 are quite similar in structure, in that the fingers in region 22, and the main pillar in region 44 are both significantly red-shifted with respect to the rim of the cloud that they are associated with.

\begin{figure}
\includegraphics[width=\columnwidth]{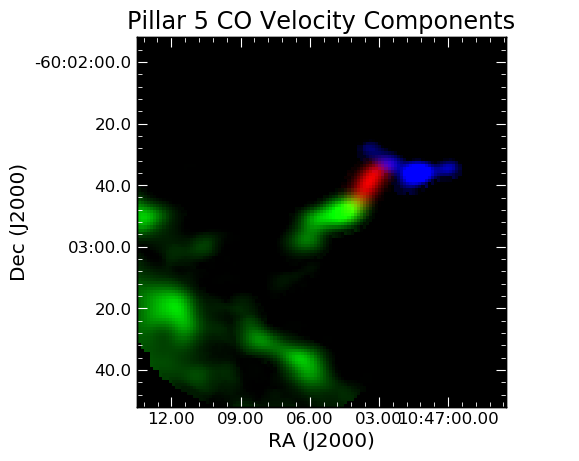}
\caption{Individual line of sight velocity components in the Pillar 5 region which show that there are three velocity separated components along the line of sight. The velocity ranges used for R,G and B were (-13.4,-8.1), (-26.8,-17.0), and (-31, -26.8) km s$^{-1}$ (respectively)}
\label{fig:pillar5_components}
\end{figure}

For Pillar 5, there are a number of distinct velocity components along the line of sight that align to appear as though they are a single pillar in the {\it Spitzer} 8 $\mu$m emission map. This is demonstrated further in Figure \ref{fig:pillar5_components}, where the integrated intensities of the various CO components seen in the spectrum are plotted as independent colours in the RGB image, with each colour representing emission in the velocity ranges of (-13.4,-8.1), (-26.8,-17.0), and (-31, -26.8) km s$^{-1}$ for R, G, and B, respectively.  That there is an $\sim$ 11 km s$^{-1}$ velocity difference between the cloud rim and the tip of the main body of the pillar (the green and the red emission, respectively) suggests that these two features are indeed separated along the line of sight, and only appear connected in projection.  Similarly, the blue shifted emission that appears to be coming from the tip of the pillar (the blue and red components in Figure \ref{fig:pillar5_components}), is separated by roughly 19 km s$^{-1}$ from the red shifted emission, further suggesting a second chance alignment along the line of sight.  It is only through analyses such as these, with the velocity information of the pillar structures, that we are able to distinguish these types of chance alignments.

Similarly, the velocity structures seen in region 45 (see Figure \ref{fig:pillar45}) show that there are again, likely two pillars along the line of sight, with the smaller one redshifted from the larger one. The MUSE data corroborate this conjecture, with distinct silhouettes observable in the H$\alpha$ emission (see Figure \ref{fig:pillar45_MUSE}).

\begin{figure}
\includegraphics[width=\columnwidth]{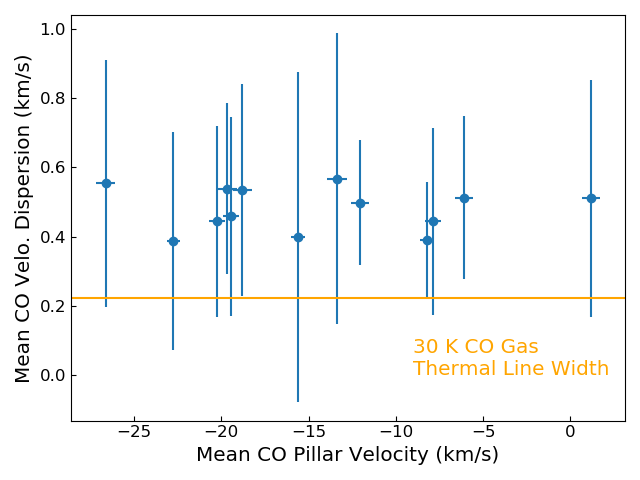}
\caption{Averaged CO velocity dispersions in each pillar. The x error bars are the same as in Figure \ref{fig:internal_motion}, and the y error bars represent the interquartile range in the velocity dispersions across each pillar.}
\label{fig:vel_dispersions}
\end{figure}

\subsection{Virial Stability}

Figure \ref{fig:vel_dispersions} shows the mean velocity dispersion of the CO for each of the pillars.  In all cases, the CO emission has a low velocity dispersion, but remaining higher than what is expected for a thermally supported line at T = 30 K. This suggests that there are support mechanisms for the gas beyond thermal.

Our sensitivity limits suggest we would be able to detect 30 K gas in our beam at levels below 0.001 M$_\odot$ in CO if the emission fully filled the beam, and the densest cores in each pillar can be traced by the C$^{18}$O emission, when present.  Where C$^{18}$O is detected, we performed a virial analysis of the emission, and find that most cores are unstable against virial collapse (see Table \ref{tab:virial}). This is expected from regions emitting in C$^{18}$O, some of which show evidence for protostellar jets \citep{Smith10_HST,Reiter16}, and outflows (this work, and Reiter et al. in prep).

\begin{center}
\begin{table}
\centering
\caption{C$^{18}$O Derived CO emission peak Masses, Virial Masses, and Stability Estimates. Masses given in M$_\odot$.}
\begin{tabular}{rrrrrc}
\hline \hline
Pillar & C$^{18}$O& FWHM &\multicolumn{2}{c}{Mass (10$^{-3}$ M$_\odot$)} & Virial\\

Num. & Num. & (km s$^{-1}$) &Core & Virial& Stable \\
\hline
2 & 1 & 0.6 & 52.8 & 7.6 & N \\
3 & 1 & 0.6 & 2.0 & 3.0 & Y \\
3 & 2 & 0.6 & 3.0 & 3.8 & Y \\
3 & 3 & 0.6 & 1.9 & 3.1 & Y \\
4 & 1 & 0.7 & 6.9 & 7.5 & Y \\
4 & 2 & 0.7 & 1.7 & 4.8 & Y \\
4 & 3 & 0.7 & 0.6 & 2.8 & Y \\
5 & 1 & 0.5 & 4.9 & 3.1 & N \\
6 & 1 & 0.6 & 8.3 & 3.5 & N \\
6 & 2 & 0.6 & 3.5 & 3.0 & N \\
8 & 1 & 0.8 & 12.0 & 13.2 & Y \\
8 & 2 & 0.8 & 0.4 & 3.3 & Y \\
16 & 1 & 0.8 & 161.0 & 21.7 & N \\
16 & 2 & 0.8 & 7.3 & 7.4 & Y \\
17 & 1 & 0.5 & 60.3 & 6.3 & N \\
18 & 1 & 1.0 & 3.2 & 6.3 & Y \\
18 & 2 & 1.0 & 3.7 & 6.9 & Y \\
18 & 3 & 1.0 & 12.8 & 12.8 & N \\
18 & 4 & 1.0 & 2.6 & 6.3 & Y \\
20 & 1 & 0.5 & 5.0 & 2.4 & N \\
22 & 1 & 0.4 & 135.8 & 4.1 & N \\
22 & 2 & 0.4 & 2.8 & 0.8 & N \\
44 & 1 & 0.6 & 2.7 & 2.8 & Y \\
44 & 2 & 0.6 & 2.3 & 3.2 & Y \\
44 & 3 & 0.6 & 1.3 & 2.6 & Y \\
45 & 1 & 0.3 & 1.0 & 0.7 & N \\
45 & 2 & 0.3 & 1.5 & 0.8 & N \\
45 & 3 & 0.3 & 2.0 & 1.0 & N \\
\hline
\end{tabular}
\label{tab:virial}
\end{table}
\end{center}

\subsection{Individual Pillar Properties}

\subsubsection{Pillar 2}

As shown in Figure \ref{fig:pillar2}, there appears to be a protostar towards the tip of this pillar. Indeed, there is a candidate Herbig-Haro (HH) object coming from this same region \citep{Smith10_HST,Reiter16}.  The three colour CO image shows a clear edge of C$^{18}$O emission (in blue) on the right side of the top half of the pillar, suggesting that perhaps the ionisation conditions are greater there, which is causing the C$^{18}$O to be brighter than the generally more abundant species.   There is an additional HH jet (HH 903) emerging from an embedded protostar located toward the middle of the pillar\citep{Ohlendorf2012}, just above the curve, however the molecular gas does not show any outflow motions associated with the jet, suggesting the jet has un-coupled from the embedded material.

\subsubsection{Pillar 3}

At first glance, the molecular gas emission in Pillar 3 appears to be quite sparse, and perhaps missing some large scale structure, which is causing the mottled appearance of the emission. However CO and $^{13}$CO are both very well detected in this region, and the mottled emission corresponds to the clumpiness seen in the {\it Spitzer} 8$\mu$m emission shown in the top left panel of Figure \ref{fig:pillar3}.  One of the longest and most powerful jets in Carina, HH~666, is at the tip of this pillar \citep{Smith2004,Smith10_HST,Reiter2015_hh666}.

The CO first moment map shown in the bottom left panel of Figure \ref{fig:pillar3} shows significantly red-shifted material to the left of the protostar. What is not as apparent in this figure is that there is a significant blue shifted tail to the CO emission on the opposite side of the protostar. It is likely there is an outflow at this pillar tip which is only somewhat spatially resolved by our observations.

\subsubsection{Pillar 4}

As with the above pillars, this region has what appears to be a protostar at its tip \citep{Ohlendorf2012}, with an associated candidate HH object, and bright rim of H$\alpha$ emission \citep{Smith10_HST}. The velocity dispersion of the CO at the location of this protostar is an order of magnitude greater ($\sim$ 5 km s$^{-1}$) than that of the rest of the pillar ($\sim$ 0.5 km s$^{-1}$). This sudden jump in the velocity dispersion suggests that there is a potential line of sight molecular outflow from this protostar.

This pillar most resembles the one presented in Figure 6 of \citet{Gritschneder10}, where a protostellar core is separating from its pillar, as we suggest is also the case here given the probable outflows coming from the core at the pillar tip.

\subsubsection{Pillar 5}

The dust emission from this region suggested a somewhat narrow pillar protruding from a rim of emission, headed by a dusty bar of emission which could have been tracing some outflow or dusty jet.  However, with the velocity information for this region, we see that there are at least two, and likely three, (velocity) distinct emission regions along the line of sight that are contributing to this apparently single pillar.  These components can be seen in green, red and blue (respectively from left to right, in Figure  \ref{fig:pillar5_components}), and are independent of the emission ridge seen in orange and yellow in the bottom left corner of the moment one map in Figure \ref{fig:pillar5}.

\subsubsection{Pillar 6}

Multiple HH objects surround pillar 6 \citep{Smith10_HST}, apparently tracing at least four distinct protostellar jets \citep{Reiter16}. Candidate jet-driving protostars for two of these were identified by \citet{Ohlendorf2012}. \citet{Reiter16,Reiter16b} demonstrated that another jet emerges from a relatively unobscured candi1 date young stellar object \citep[using the catalog from][]{Povich2011}. 

There do not appear to be any  molecular outflows associated with the HH jets. It is interesting to note that these star forming cores appear to be separated by Jeans lengths, and, that they have a different  mean velocity ($\sim$ -12 km s$^{-1}$) than the less dense pillar material between them ($\sim$ -13 km s$^{-1}$), which is significant given that the velocity resolution of these observations is 0.08 km s$^{-1}$.

\subsubsection{Pillar 8}

Pillar 8 contains the Treasure Chest \citep[cf][]{smith2005} which is host to a proto-cluster rather than individually forming stars. Its most interesting kinematic structure is the velocity gradient along the `straight edge' to the left of images in Figure \ref{fig:pillar8}.  The sharp straight line corresponding to the Eastern edge of this pillar appears to be a single coherent structure with a velocity gradient indicating that the bottom portion of the pillar edge is red-shifted with respect to the main body of the observed emission. Because the exact orientation of the overall pillar (with respect to the line of sight) is unknown, it is not clear whether this velocity gradient is indicative of red-shifted material flowing into or out of the pillar.

The mass we derive from our CO observations for this region is about twice that derived by \citet{Roccatagliata13} from the {\it Herschel} observations of this same region.

\subsubsection{Pillar 16, 17, 18 complex}

Pillars 16, 17, and 18 were observed in three separate mosaics, and while combined into a single large map for imaging, were analysed individually. The substructure present in Pillar 16 (central pillar) is much more intricate in our CO observations than expected from the {\it Spitzer} 8 $\mu$m emission, suggesting, like in other regions, that there are multiple components along the line of sight contributing to what looks like a single pillar in the dust continuum.  Pillar 17 (leftmost pillar) shows a velocity gradient across it unlike any seen in Pillars 16 or 18, but similar to that seen in Pillar 2, while Pillar 18 (rightmost pillar) shows evidence for redshifted emission towards the central portions of the pillar. Both Pillars 17 and 18 show an excess of C$^{18}$O emission at their edges (with a much smoother emission profile in Pillar 17) than would be expected from a generalised abundance ratio.

\subsubsection{Pillar 20}

From our largest individual pillar map, the pillar shown in Figure \ref{fig:pillar20} forms part of the dust rim surrounding Trumpler 15. The relationship between this and Pillar 22, towards the other end of the rim is discussed in the section below.  Here we note that this Pillar has a coherent velocity structure from its tip to its core, yet it appears that where the pillar joins with the ridge of emission surrounding Trumpler 15, the velocities diverge, depending on which side of the pillar the gas is on.  It is clear however, that all three CO isotopologues are only tracing the outer edges of this pillar. The C$^{18}$O is only detected towards the tip of the pillar, while the $^{12}$CO and $^{13}$CO are mostly found on the pillar edges.  With the inclusion of the total power data, it is unlikely that this effect is due to missing large scale structures, but is instead more likely to be due to density and/or temperature effects that cause the CO be less abundant in the centre of this pillar.

The eastward spur coming from the tip of the pillar does show some signs for red-shifted emission, which could be indicative of an outflow. 
\citet{Ohlendorf2012} identified a compact green object (CGO) at the tip of this pillar. The point-like source appears green in three colour images made with \textit{Spitzer} data because of excess emission in the 4.5~\micron\ band, possibly tracing shock-excited H$_2$ emission in a protostellar outflow. 

\subsubsection{Pillar 22}

The region observed here shows the opposite edge of the rim also explored in the section above (Pillar 20).  It was chosen because it appears to have a number of small pillars which are still attached to the larger rim structure.  From the first moment map of the CO emission in this region (bottom left panel of Figure \ref{fig:pillar22}), it appears that those fingers are velocity shifted (blue shifted) from the rim, however the transition from pillar to rim velocities is smooth, suggesting that a velocity gradient in a coherent structure, rather than multiple line of sight components (like seen in Pillar 5).  The pillar at approximately (0, -25) arcsec offset from zero in Figure \ref{fig:pillar22} has much stronger C$^{18}$O emission, as shown in the RGB map in the top right corner of the figure. 

\subsubsection{Pillar 44}

The pillar imaged in this region has a very distinct velocity ($\sim$ -10 \kms{}) from that of the cloud rim it is associated with ($\sim$ -7.5 \kms{}), with little velocity variation within it.  The spectrum, shown in the bottom right corner of Figure \ref{fig:pillar44} shows a brightest single peak for each CO isotopologue, with an additionally blueshifted component in the $^{12}$CO. This appears to be foreground/background diffuse emission towards the left portion of the map. The velocity offsets between the dominant peaks of the $^{13}$CO and C$^{18}$O emission in the spectrum of this pillar come from the differering spatial distributions of the two isotopologues.  When averaged over small portions of the pillar (for instance, the blueish region in the CO RGB image), the spectra are consistent in their velocity profiles, however the $^{13}$CO is brighter in different portions of the pillar than the C$^{18}$O, causing the overall appearance of a velocity discrepancy between the species.

This pillar was also observed by \citet{McLeod16}, who found a jet coming from the pillar tip, indicative of star formation.

\subsubsection{Pillar 45}

This structure, as shown in Figures \ref{fig:pillar45_MUSE} and \ref{fig:pillar45}, appears to be at least two pillars superposed along the line of sight at distinct velocities. The main component is the largest scale structure, identifiable as emitting in the -4 to -7 \kms{} velocity range, with a second, smaller component at velocities more blueshifted than -10 \kms{}.  The C$^{18}$O in this region appears to be concentrated at the tips of the two pillars, indicating they are likely being similarly irradiated.

\section{Discussion}
\label{sec:discussion}

\subsection{Comparisons between pillar structures}
\label{sec:pillar_comparison}

The pillar velocity structures are, by and large, consistent with the structures seen in the lower resolution CO (J=1-0) observations of \citet{Rebolledo16}, who showed single peaked velocity profiles in their Region 1, which roughly corresponds to our Pillar 2, and multiple velocity components towards their Region 2. This is a similar profile to that seen in our regions 16-22, which are in and around their Region 2 (compare the bottom right panels of Figures \ref{fig:pillar2} - \ref{fig:pillar45} to the leftmost panels of their Figure 4). Their Region 3 was outside of our sampled region.

In Pillars 22 and 44 (shown in Figures \ref{fig:pillar22} and \ref{fig:pillar44}), we see that the protruding pillar (or fingers in the case of region 22) are at distinct velocities from the cloud edges that they are associated with.  These are the only two of those we specifically observed with  cloud edges that show these velocity offsets, suggesting that these offsets are rare occurrences.  The velocity structures seen in these two regions are different from those seen in Pillar 5 which appears to have a few independent cores/pillars along the line of sight. 

That the velocity structures of each finger in the Pillar 22 region are virtually identical suggests fingers that are being pushed, or photo-ablated into the cloud rim.  We note that Pillars 22 and 20 appear, from the Herschel image in Figure \ref{fig:context}, to be attached to the same cloud rim surrounding Trumpler 15.  There is no information about the 3D structure of this rim surrounding the star cluster, however, the velocities of the gas are similar on the two sides of the cluster: the upper portion of region 20 has velocities of roughly -10 km s$^{-1}$, which is the same as the velocity of the 'red-shifted' fingers of Pillar 22.  There is a velocity difference between the sampled cloud rim components, which suggests some bulk motion of the bubble (e.g. expansion), or that we are seeing one edge more in projection than the other. If we assume the cloud is expanding, given that the velocity difference between the cloud edge in Pillar 20 (-10 km s$^{-1}$) and Pillar 22 (-6 km s$^{-1}$) is 4 km s$^{-1}$, and that the projected separation is 0.12 pc, the cloud has been expanding for roughly 30 000 years, a much shorter timescale than the age of the cluster embedded within it \citep[see the discussion in ][]{Hanes18}.

\begin{figure}
\includegraphics[width=\columnwidth]{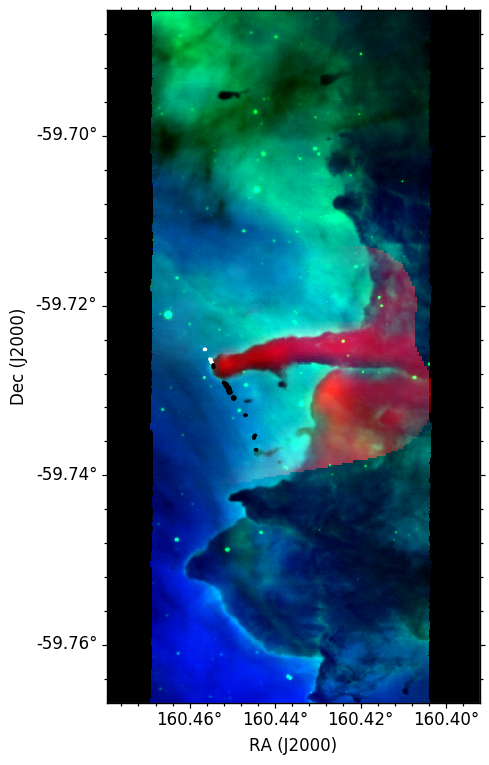}
\caption{Pillar 44 as observed with MUSE in H$\alpha$ (blue) and [SII] (green), with the ALMA CO emission plotted in red.}
\label{fig:pillar44_MUSE}
\end{figure}

Whether the blueshifted emission in Pillar 44 is physically associated with its cloud edge is less clear. The analysis of the optical lines observed with MUSE in \citep{McLeod16} shows similar irradiation of the pillar and its cloud edge, however their higher resolution observations (overplotted with red ALMA CO observations in Figure \ref{fig:pillar44_MUSE}) do suggest a slight distinction between the blueshifted pillar and the cloud edge, and towards the base of the pillar, they appear to come together.

Interestingly, neither Pillars 2 or 4, show any of these velocity distinctions from their cloud rims. These two pillars are more similar to the traditional picture of a pillar being photo-ionised by its surroundings. Unlike those in Pillar 22 (but similarly to Pillar 44), these pillars do show evidence for star formation in the form of protostellar jets \citep{Reiter16}. Further followup on these pillars at higher resolution, and with better tracers of the photon dominated regions at the edges of these pillars would give a better understanding of why Pillars 22 and 44 have velocity offsets while 2 and 4 do not. Is it an evolutionary effect? Or is this more strongly tied to the method of photo-ionisation, or wind forces being felt by the pillars?

The velocities of the gas in Pillar 6, which is detached from any cloud, are also interesting. In this region, the cores (as traced by C$^{18}$O), are separated by Jeans lengths. In between the cores, the gas appears to be more red-shifted than the cores themselves (by roughly 1 km s$^{-1}$, well resolved spectrally by our observations).  The H$\alpha$ emission from this region \citep[as observed by][]{Smith10_HST}, is brightest along the right edge of the pillar, and not towards its tip as expected from a model of a pillar being reveled through photo-ionisation.  If there is a nearby star that is photo-ionising Pillar 6 from the right, it appears to be able to push the lower density inter-core material more easily than the higher density cores themselves.  However, Figure \ref{fig:context} does not show any O or B stars close enough to Pillar 6, at the right orientation (and right radial velocity) to be causing this H$\alpha$ emission and velocity discrepancy.

The tips of Pillars 3, 4, 45, and left most and horizontal pillars in the central region of the Pillar 16, 17, 18 mosaic all show 'blue' tips in the CO isotopologue RGB images. That the C$^{18}$O emission is anomalously bright at this pillars tips does suggest that irradiation may be playing a greater role these locations than the rest of each of these pillars - because C$^{18}$O is more excitable than the other two isotopologues. This is further supported by the intense C$^{18}$O emission at the right edge of Pillar 2 (see the top left panel of Figure \ref{fig:pillar2}), which is also extremely bright in H$\alpha$, as discussed in \citet{Smith10_HST}. Indeed, for all the pillars with 'blue' regions in the RGB images (e.g. an over abundance of C$^{18}$O with respect to CO and $^{13}$CO), there is a strong correlation with the presence of H$\alpha$.

Pillar 8 is known to contain significant star formation in its core, and so any record of its previous kinematics will necessarily be confused with those motions.  It appears to have a generally uniform velocity structure, however, as seen in Figure \ref{fig:pillar8}, the strikingly straight edge on its left side does appear to have two velocity components. This is suggestive of a blue shifted ridge that is being pushed by the more red-shifted star forming region. Further investigation of these kinematics is saved for future work.

\begin{figure}
\includegraphics[width=\columnwidth]{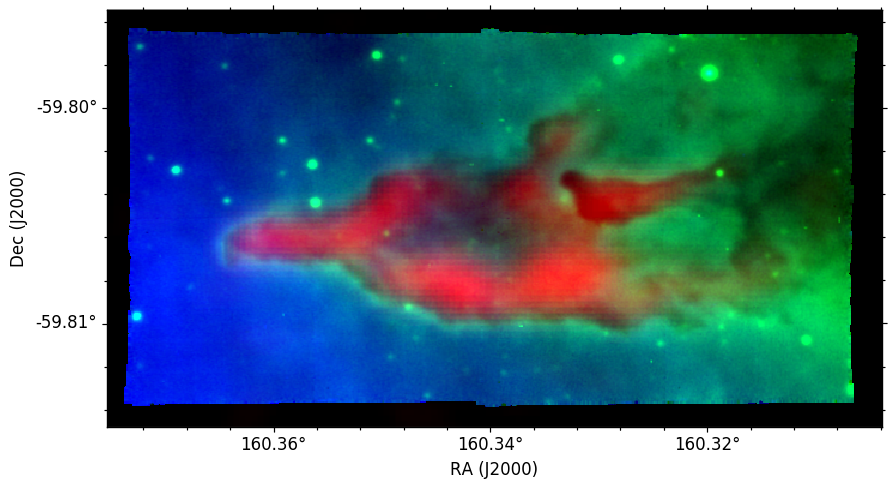}
\caption{Pillar 45 as observed with MUSE in H$\alpha$ (blue) and [SII] (green), with the ALMA CO emission plotted in red.}
\label{fig:pillar45_MUSE}
\end{figure}

The CO emission for Pillars 44 and 45 are compared to the H$\alpha$ and [S{\sc ii}] emission from these same regions observed with MUSE \citep[and presented in][]{McLeod16} in Figures \ref{fig:pillar44_MUSE} and \ref{fig:pillar45_MUSE}. Here we see that the CO emission from ALMA is brightest where the optical tracers are darker. In comparing the emission in  Figure \ref{fig:pillar45_MUSE} to the velocity information in Figure \ref{fig:pillar45} we see that what appeared as a single pillar in the MUSE data appears as two components in the ALMA data with a   velocity distinct and blue shifted CO emission component towards the right side of these figures.

As described in Section \ref{sec:MN}, each of the pillars observed for this study has peak column densities high enough to form high-mass stars (i.e. above a threshold of 10$^{21}$ cm$^{-2}$).  There seems to be a general trend that those with higher ($\gtrsim4\times10^{21}$ cm$^{-2}$) peak column densities are more likely to show signposts of ongoing star formation (i.e. jets in optical tracers), but beyond that, there does not appear to be any trend with star formation efficiency.  Pillar 6 appears to be the most prolific star forming pillar, with signposts  at roughly every Jeans length, while the rest, mostly with higher peak column densities have much lower star formation efficiencies. This is likely a reflection of higher peak densities in Pillar 6 in the past. If Pillar 6 had a higher density, and higher mass loss rate early on in its evolution, then that could explain its currently high star formation efficiency in comparison to its current (relatively) lower peak column density. This pillar is extremely bright in H$\alpha$ (Reiter et al. in prep) which is indicative of a high incident radiation field which could have triggered high mass-loss rates.

\subsection{Comparisons to Models}

One of the key reasons for quantifying the dynamics of the observed pillars in Carina was to  directly compare their properties to those of pillars generated in simulations of star-forming regions.   Carina is large enough to sample a large range of stellar masses, and therefore we can only compare our results to those simulations with comparably large cloud masses (e.g. those above 10$^4$ M$_\odot$).  The models of \citet{Dale12II}, and \citet{Gritschneder09,Gritschneder10}, were of regions with sufficient mass to be able to draw specific comparisons, and each provided specific and measureable quantities for these comparisons. These include velocity dispersions, (potential) velocity offsets or gradients between the pillar and the parent cloud (e.g. ordered flows) and the presence of internal flows.  The predictions of these models are summarised in Table \ref{tab:models}.

As was shown in Section \ref{sec:velocities}, all of the pillars in our study have very low velocity dispersions (see also Figure \ref{fig:vel_dispersions}). There is evidence for some pillars having ordered flows, but the incidence rate is roughly 50\%, and there is no evidence for internal flows (see Figure \ref{fig:internal_motion}).  These properties are most aligned with the predictions of \citet{Dale12II}.

That we can distinguish between models which only differ in their treatment of turbulence and their initial bulk densities suggest that these are/were important initial conditions in highly irradiated and turbulent regions like Carina, and likely play a larger role than other factors such as the magnetic field.

Because the largest difference between these sets of models was the treatment of the turbulent decay spectrum, we suggest that the models of \citeauthor{Dale12II} have a better treatment of this decay, which, in their case, followed a Kolmogorov slope. 

\begin{table}
\caption{Summary of physical property predictions from models that can be directly tested by the pillar observations presented here.}
\begin{tabular}{l|ccc}
\hline\hline
Model & Velocity & Internal & Ordered\\
  & Dispersion & Flows & Motions\\
 \hline
\citet{Dale12II} & $\sim$ 1 km s$^{-1}$ & no & some\\
\citet{Gritschneder10} & $2-6$ km s$^{-1}$ & yes & none\\
\citet{Gritschneder09} & 1-2 km s$^{-1}$ & yes & none\\
\hline
\end{tabular}
\label{tab:models}
\end{table}

\section{Conclusions}
\label{sec:conclusions}

We observed 13 pillars in the Carina nebula in CO J=2-1 and its isotopologues.  We find pillar masses of order 30-2000 M$_\odot$ depending on the size of the pillar, and mean column densities of order a few $\times$ 10$^{20}$ cm$^{-2}$.  Where we have detected C$^{18}$O, we have performed virial analyses and find that most embedded cores are virial unstable, an likely to collapse.  On average, we also find that most cores within these pillars are separated by 1-2 jeans lengths. We find interesting velocity structures in these pillars, and have compared those structures to predictions from star formation models which include photo-ionisation. We find low velocity dispersions ($<1$ km s$^{-1}$), no internal motions (by comparing the CO to the $^{13}$CO mean velocities), and that some of the pillars do show ordered flow motions, but not all. These results appear to be most consistent with the predictions presented in the models of \citet{Dale12II}.

Further study of the molecular gas kinematics in these pillars will give greater insights into the current generation of star formation in Carina through studies of outflow detection and classification of properties and comparison to known jets observed at optical wavelengths.  Additional [C{\sc I}] (from both MUSE and ALMA) can also be used with this data and existing H$_2$ and H$\alpha$ observations to place constraints on the photon dominated regions (PDRs) at the edges of these pillars, as has been done in Orion by \citet{Goicoechea16}, in anticipation of being able to use JWST to quantify the PDRs at high resolution with MIRI and NIRSpec.

\section*{Acknowledgements}

JCM acknowledges support from the European Research Council under the European Community's Horizon 2020 framework program (2014-2020) via the ERC Consolidator grant `From Cloud to Star Formation (CSF)' (project number 648505).
MR acknowledge funding from the European Union's Horizon 2020 research and innovation programme under the Marie Sk\'{l}odoska-Curie grant agreement No.\ 665593 awarded to the Science and Technology Facilities Council.
The authors would like to thank Daniel Seifried and Matthew Povich for helpful discussions during the writing of this manuscript.
This research has made use of NASA's Astrophysics Data System.  This research has made use of the SIMBAD database, operated at CDS, Strasbourg, France. This work made use of the IPython package \citep{ipython} This research made use of APLpy, an open-source plotting package for Python hosted at http://aplpy.github.com This research made use of NumPy \citep{numpy} This research made use of matplotlib, a Python library for publication quality graphics \citep{matplotlib} This research made use of Astropy, a community-developed core Python package for Astronomy \citep{astropy} This research made use of spectralcube, a library for astronomical spectral data cubes \citep{spectral_cube}. This research made use of pyspeckit, an open-source spectral analysis and plotting package for Python hosted at http://pyspeckit.bitbucket.org  This paper makes use of the following ALMA data: ADS/JAO.ALMA\#2016.1.00101.S. ALMA is a partnership of ESO (representing its member states), NSF (USA) and NINS (Japan), together with NRC (Canada) and NSC and ASIAA (Taiwan), in cooperation with the Republic of Chile. The Joint ALMA Observatory is operated by ESO, AUI/NRAO and NAOJ. The acknowledgements were compiled using the Astronomy Acknowledgement Generator.

\appendix

\section{CO Moment Maps}

Here we present the CO moment maps for each pillar. In each figure, the left panel shows a 3 colour (RGB) image of the CO, $^{13}$CO and C$^{18}$O (respectively) emission from each mapped region.  The right panel shows the intensity weighted velocity (moment one) map for the CO emission in that region.

\foreach \x in {2,3,4,5,6,8}
{	\begin{figure*}
		\includegraphics[width=0.9\textwidth]{Figures/Pillar_\x_4panel_full_page.png}
    	\caption{Emission from the Pillar \x{} region. {\bf Top left:} Spitzer 8 $\mu$m emission taken from the IPAC archive overlaid with the CO integrated intensity contour that corresponds to the `threshold for star formation' as described in the text. {\bf Top right:} RGB image of the integrated intensities of CO, $^{13}$CO and C$^{18}$O (respectively) in the observed region. {\bf Bottom Left:} Intensity weighted velocity map of the CO in the observed region, {\bf Bottom Right:} Spectra of CO, $^{13}$CO and C$^{18}$O (in red, green and blue, respectively) integrated over the observed region.      }
    \label{fig:pillar\x}
    \end{figure*}
}

\begin{figure*}
		\includegraphics[width=0.9\textwidth]{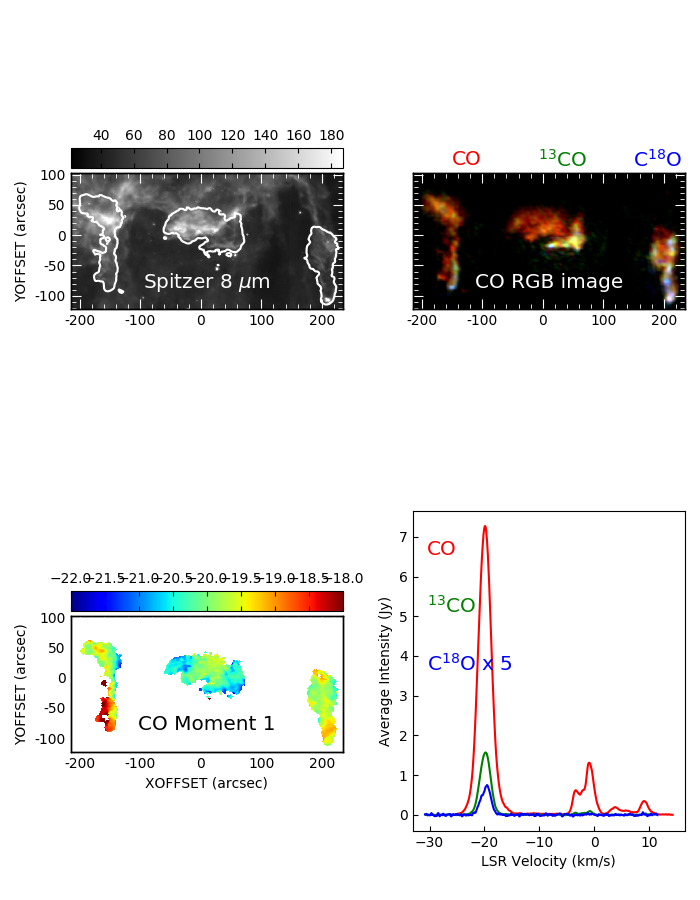}
    	\caption{Emission from the Pillar 16, 17 and 18 region. {\bf Top left:} Spitzer 8 $\mu$m emission taken from the IPAC archive overlaid with the CO integrated intensity contour that corresponds to the `threshold for star formation' as described in the text. {\bf Top right:} RGB image of the integrated intensities of CO, $^{13}$CO and C$^{18}$O (respectively) in the observed region. {\bf Bottom Left:} Intensity weighted velocity map of the CO in the observed region, {\bf Bottom Right:} Spectra of CO, $^{13}$CO and C$^{18}$O (in red, green and blue, respectively) integrated over the observed region.      }
\label{fig:pillar161718}
\end{figure*}

\foreach \x in {20,22,44,45}
{	\begin{figure*}
		\includegraphics[width=0.9\textwidth]{Figures/Pillar_\x_4panel_full_page.png}
    	\caption{Emission from the Pillar \x{} region. {\bf Top left:} Spitzer 8 $\mu$m emission taken from the IPAC archive overlaid with the CO integrated intensity contour that corresponds to the `threshold for star formation' as described in the text. {\bf Top right:} RGB image of the integrated intensities of CO, $^{13}$CO and C$^{18}$O (respectively) in the observed region. {\bf Bottom Left:} Intensity weighted velocity map of the CO in the observed region, {\bf Bottom Right:} Spectra of CO, $^{13}$CO and C$^{18}$O (in red, green and blue, respectively) integrated over the observed region.      }
    \label{fig:pillar\x}
    \end{figure*}
}

\bibliographystyle{mnras}
\bibliography{mnras_template}

\bsp	
\label{lastpage}
\end{document}

%% file: obs_params.tab
\begin{table*}
\caption{Calibrators and Observing Parameters for ALMA project 2016.1.00101.S.}
\begin{tabular}{lcclllcccc}
\hline \hline
Pillar&Time on& \# 7m & \multicolumn{3}{c}{Calibrators} & & \multicolumn{2}{c}{Baseline Lengths} & PWV\\ \cline{4-6} \cline{8-9}
Number  & Source & Mosaics & Phase & Bandpass & Flux & Observing Date & Max & Min &\\
 & (sec) &Pointings &&&&&(m) &(m) & (mm)\\
 \hline 

Pillar 2 & 170 & 53 &J1147-6753 & J0522-3627 & J0522-3627 & 2016-11-22 & 45 & 8.9 &0.82\\
Pillar 3 & 212 & 95 &J1147-6753 & J1256-0547 & J1058+0133 & 2016-10-27 & 48 & 8.9 &0.53\\
Pillar 4 & 191 & 57 &J1147-6753 & J0522-3627 & J0522-3627 & 2016-11-22 & 45 & 8.9 &0.82\\
Pillar 5 & 33 & 18 &J1147-6753 & J0522-3627 & J0522-3627 & 2016-11-26 & 45 & 8.9 &0.44\\
Pillar 6 & 11 & 22 & J1147-6753 & J0522-3627 & J0522-3627 & 2016-11-26 & 45 & 8.9 &0.44\\
Pillar 8 & 265 & 115 &J1147-6753 & J1058+0133 & J1058+0133 & 2016-11-26 & 45 & 8.9 &0.42\\
Pillar 20 & 382 & 149 &J1147-6753 & J1058+0133 & J1058+0133 & 2016-11-27 & 45 & 8.9 &1.22\\
Pillar 22 & 74 & 30 &J0940-6107 & J1058+0133 & J1058+0133 & 2016-10-31 & 45 & 8.9 &0.48\\
Pillar 44 & 21 & 11 &J0940-6107 & J1058+0133 & J1058+0133 & 2016-10-31 & 45 & 8.9 &0.48\\
Pillar 45 & 149 & 59 &J0940-6107 & J1058+0133 & J1058+0133 & 2016-10-31 & 45 & 8.9 &0.48\\
Pillar 16 & 106 & 30 &J1147-6753 & J1058+0133 & J1058+0133 & 2016-10-28 & 45 & 8.9 &0.38\\
Pillar 17 & 106 & 30 &J1147-6753 & J1058+0133 & J1058+0133 & 2016-10-28 & 45 & 8.9 &0.38\\
Pillar 18 & 138 & 38 & J1147-6753 & J1058+0133 & J1058+0133 & 2016-10-28 & 45 & 8.9 &0.38\\
 
 \hline \hline
\label{tab:obs_params}
\end{tabular}
\end{table*}